\documentclass{article}
\usepackage[letterpaper, portrait, margin=1in]{geometry}
\usepackage[utf8]{inputenc}
\usepackage{graphicx, caption}
\usepackage{amsfonts}
\usepackage{amsmath,bm}
\usepackage{cite}
\usepackage{IEEEtrantools}
\usepackage{amssymb}
\usepackage{amsthm}
\usepackage{authblk}
\usepackage{color}
\usepackage{comment}
\usepackage{mathtools}
\usepackage{braket}
\usepackage{enumitem}
\usepackage{multirow}
\usepackage{multicol}
\usepackage{hyperref}
\usepackage{balance}
\usepackage{flushend}
\usepackage{tikz}
\usepackage{subcaption}
\usepackage{adjustbox}
\usepackage{pgfplots}
\usetikzlibrary{quantikz}

\newcommand{\Tof}{\textnormal{Tof}}

\theoremstyle{plain}
\newtheorem{thm}{Theorem}

\newtheorem{lemma}{Lemma}
\newtheorem{corollary}{Corollary}

\newtheorem*{remark}{Remark}

\title{Almost-Optimal Computational Basis State Transpositions}
\author[1,2]{Steven Herbert}
\author[1]{Julien Sorci}
\author[1]{Yao Tang}
\affil[1]{\small \textit{Quantinuum, Terrington House, 13–15 Hills Road, Cambridge CB2 1NL, United Kingdom}}
\affil[2]{\small \textit{Department of Computer Science and Technology, University of Cambridge, United Kingdom}}

\begin{document}

\maketitle

\begin{abstract}
    We give an explicit construction to perform any $n$-qubit computational basis state transposition using $\Theta(n)$ gates. This nearly coincides with the lower bound $\Omega(n/\log(nd))$ on worst-case and average-case gate complexity to perform transpositions using a $d$-element gate-set, which we also prove.
\end{abstract}

\section{Introduction}

Quantum circuits that permute computational basis states are widely found in quantum computing: the $X$, CNOT and Toffoli gates do exactly this, and blocks of $\{ X, \textnormal{CNOT}, \textnormal{Toffoli} \}$ are found, for example, every time an oracle is invoked to compute a classical function. Indeed, owing to the quantum \textit{computational} universality of the gate-set $\{ H, \textnormal{Toffoli} \}$ \cite{Aharonov2003}, every quantum circuit can be replaced by a functionally equivalent version represented as alternating blocks of permutations and Hadamard gates. Furthermore, it has been observed that many of the most powerful quantum circuits amount to no more than a computational basis state permutation conjugated by a transform, such as the Fourier or Schur transform. \cite{Hav19}. Some notable instances featuring permutation circuits are Shor's factoring algorithm, which employs a permutation circuit for modular exponentiation, and the discrete-time quantum walk operator, which acts as a diffusion operator followed by a conditional shift -- the latter of which being a permutation circuit \cite{Shor1994, Kempe2003}.

\makeatletter{\renewcommand*{\@makefnmark}{}
\footnotetext{$^*$ Equal contributions: author order is alphabetical \\ Contact: Steven.Herbert@quantinuum.com, Julien.Sorci@quantinuum.com, Yao.Tang@quantinuum.com}\makeatother}

Owing to the general importance of permutations in quantum circuits, we explore bounds on performing arbitrary computational basis state \textit{transpositions}. Specifically, we consider an $n$-qubit circuit with computational basis states $\{ \ket{x} : x \in \{0,1 \}^n\}$, and are interested in the gate complexity of the operation:
\begin{equation}
\label{transpositionmap}
    \ket{x} \mapsto \begin{cases} \ket{x}, & \textnormal{if } x \notin \{ a , b \} \\ 
                                  \ket{b}, & \textnormal{if } x = a \\
                                  \ket{a}, & \textnormal{if } x = b \\ 
    \end{cases}   
\end{equation}
for fixed but arbitrary $a, b \in \{0,1\}^n$. 

As a concrete example of an important quantum circuit primitive where computational basis transpositions manifest, consider a quantum oracle which evaluates a classical function $f:\{0,1\}^n \rightarrow \{0,1\}^m$. This acts on the computational basis states as the mapping $\ket{x,y} \mapsto \ket{x, y \oplus f(x)}$ for $x \in \{0,1\}^n$ and $y \in \{0,1\}^m$, where $\oplus$ denotes bitwise addition. Such a mapping can be compiled as a product of computational basis state transpositions which transpose $\ket{x, y}$ with $\ket{x,y \oplus f(x)}$, one for each $y \in \{0,1\}^m$ and $x \in \{0,1\}^n$ such that $f(x)$ is nonzero. Compiling this circuit as a product of transpositions is particularly advantageous for functions with small support.

Previous work on the compilation of permutation circuits has largely focused on the complexity of compiling an arbitrary computational basis state permutation. The worst-case gate complexity was shown to be $\Omega(n2^n/\log(n))$ in Ref. \cite{Shende2003} and constructions which nearly meet this lower-bound have been proposed in Ref. \cite{Saeedi2010} and Ref. \cite{Li2023}. On the other hand, there appears to be little in the literature on the compilation of a computational basis state transposition. Noting that the set of transpositions generates the full group of permutations, transpositions constitute an important building block for quantum circuits in general.

The organisation of this note is as follows. In Section~\ref{sect:lower-bound} we prove a lower bound on the worst-case gate complexity to compile a unitary from a given family of unitary matrices, and show that the same asymptotic lower bound holds for the average gate complexity, independent of the number of ancilla qubits present. We specialise these results to the case of computational basis state transpositions. In Section~\ref{sect:achieve-exact} we give a construction for a circuit that performs any transposition with $\Theta(n)$ gates and either two or $n-1$ clean ancillas, which nearly achieves the lower-bound of $\Omega(n/\log(nd))$ for a $d$-element gate-set proved in the preceding section. In Section~\ref{sect:numericalresults} we present numerical results demonstrating the performance of our proposed method of performing a computational basis state transposition in terms of CNOT and T gate counts. Lastly, in Section~\ref{sect:discussion} we conclude the paper with some final remarks.  

\section{A lower-bound on the gate-complexity of computational basis state transpositions}
\label{sect:lower-bound}

In this section we prove a lower-bound on the worst-case and average-case gate complexity of a computational basis state transposition for any finite gate-set. We begin by proving a worst-case lower bound for an arbitrary set of operators, and then specialise to transpositions. For the remainder of the section we will let $\mathcal{G}$ denote a finite gate-set consisting of $d$ gates with each gate acting on at most $c$ qubits for some constant $c$.

\begin{thm}
\label{lowerbound}
Let $\mathcal G$ be a finite gate-set consisting of $d$ gates. Then for any set of unitary matrices, $\mathcal{U}$, there is an element of $\mathcal{U}$ with gate complexity
\begin{equation}
\label{lowerboundasymptotics}
    \Omega\Big(\log(|\mathcal U|)/\log(nd)\Big).
\end{equation} 
Moreover, if $|\mathcal U| \in \mathcal{O}(n^n)$ then this holds even if we permit an arbitrary number of ancilla qubits.
\end{thm}

\begin{proof}
We first show that the claimed gate complexity holds if no ancillas are present. Consider an $n$-qubit circuit that is compiled by $k$ gates of $\mathcal{G}$. Since each gate in $\mathcal{G}$ acts on at most $c$ qubits then there are at most $(n!/(n-c)!)d$ ways of applying a gate from $\mathcal{G}$ to the circuit, and therefore there are at most $\big((n!/(n-c)!)d\big)^k$ possible operations that can be achieved by a circuit with $k$ gates. If every element of $\mathcal{U}$ can be compiled by such a circuit then $k$ must be large enough so that:
\begin{equation*}
    |\mathcal{U}| \leq \Big((n!/(n-c)!)d\Big)^k
\end{equation*}
or equivalently:
\begin{equation}
\label{eqn:gate-lower-bound-1}
      k \geq \log\big(\mathcal{|U|}\big)/\log\Big((n!/(n-c)!)d\Big).
\end{equation}
Therefore there is some element in $\mathcal{U}$ that requires at least $\log\Big(|\mathcal U|\Big) / \log\Big((n!/(n-c)!)d\Big)$ gates of $\mathcal G$ to be compiled. For any positive integers $n,c$ with $c \leq n$ we have the upper-bound $(n!/(n-c)!) \leq n^c$, from which it directly follows that $\log(n!/(n-c)!) \in \Theta(\log(n))$. Thus the resulting element of $\mathcal{U}$ has gate-complexity $\Omega\big(\log(|\mathcal U|)/\log(nd)\big)$, as claimed.

Next, we consider the case where an additional $m$ ancillas are available. In particular we ask whether the lower-bound on the gate complexity can be reduced from that in (\ref{lowerboundasymptotics}). First, by the premise, we are only concerned with the case where the lower bound has been reduced from $\log\big(\mathcal{|U|}\big)/\log\Big((n!/(n-c)!)d\Big)$, and so as each gate operates on at most $c$ qubits, this means that at most some 
\begin{equation*}
    n^\prime \leq c \cdot \frac{\log(|\mathcal U|)}{\log\Big((n!/(n-c)!)d\Big)}
\end{equation*}
qubits can be involved in the circuit. The assumption that $|\mathcal U| \in \mathcal{O}(n^n)$ implies that $\log(|\mathcal U|) \in \mathcal{O}(n\log(n))$, and thus $n^\prime \in \mathcal{O}(n)$. Therefore, even if an arbitrary number of ancillas are available, we can effectively upper-bound the total number of qubits by $n^\prime$ (as the ancillas are identical). It follows that we can substitute $n^\prime$ into the denominator of the expression in (\ref{lowerboundasymptotics}), however as $n^\prime \in \mathcal{O}(n)$ the asymptotic expression does not change.
\end{proof}

We note that a similar version of Theorem \ref{lowerbound} has appeared in \cite[Lemma 8]{Shende2003}. However, our more general statement will be important for the results which follow it. We now show that the same gate complexity in Theorem~\ref{lowerbound} holds on average.

\begin{thm}
\label{avgcomplexity}
Let $\mathcal G$ be a finite gate-set consisting of $d$ gates. Then for any set of unitary matrices, $\mathcal{U}$, the average gate complexity of the elements of $\mathcal{U}$ is
\begin{equation*}
    \Omega\Big(\log(|\mathcal U|)/\log(nd)\Big).
\end{equation*} 
Moreover, if $|\mathcal U| \in \mathcal{O}(n^n)$ then this holds even if we permit an arbitrary number of ancilla qubits.
\end{thm}

\begin{proof}
If we now adapt Theorem~\ref{lowerbound} to consider $\tilde{k}$ large enough such that \textit{half} of the elements of $\mathcal{U}$ can be compiled, then we obtain:
\begin{equation*}
     \tilde{k} \geq \log\big(\mathcal{|U|} / 2\big)/\log\Big((n!/(n-c)!)d\Big)
\end{equation*}
To lower-bound the average gate complexity of compiling the elements of $\mathcal{U}$ we now lower-bound:
\begin{itemize}
    \item The at most half of the elements of $\mathcal{U}$ that have been compiled within $ \log\big(\mathcal{|U|} / 2\big)/\log\Big((n!/(n-c)!)d\Big)$ operations have consumed at least $0$ operations in their compilation.
    \item The at least half of the elements of $\mathcal{U}$ that have not been compiled within $ \log\big(\mathcal{|U|} / 2\big)2\log\Big((n!/(n-c)!)d\Big)$ operations  have each consumed at least $ \log\big(\mathcal{|U|} / 2\big)/\log\Big((n!/(n-c)!)d\Big)$ to compile.
\end{itemize}
From this we can easily obtain a lower-bound on the average gate complexity:
\begin{equation*}
    k_{\text{ave}} \geq 0.5 \times 0 + 0.5 \times \log\big(\mathcal{|U|}/2\big) / \log\Big((n!/(n-c)!)d\Big).
\end{equation*}
The claim that the average complexity holds even with an arbitrary number of ancilla qubits follows by the same reasoning presented in the proof of Theorem~\ref{lowerbound}.
\end{proof}

We now specialise Theorems~\ref{lowerbound} and \ref{avgcomplexity} to deduce the worst-case and average gate complexity of a computation basis state transposition. 

\begin{corollary}
\label{cor1}
Let $\mathcal G$ be a finite gate-set consisting of $d$ gates. Then, for any $n$-bit computational basis state, $\ket{a}$ there exists another $n$-bit computational basis state, $\ket{b}$, such that the gate complexity required to compile a transposition of $\ket{a}$ and $\ket{b}$ using the gate-set $\mathcal{G}$ is $\Omega\Big(n/ \log(nd)\Big)$. In addition, the average complexity of such a transposition is $\Omega\Big(n/ \log(nd)\Big)$. Both of these lower bounds hold even if we permit an arbitrary number of ancilla qubits.
\end{corollary}

\begin{proof}
This follows directly from Theorems~\ref{lowerbound} and \ref{avgcomplexity} by taking $\mathcal U$ to be the set of transpositions with $\ket{a}$. This set has $2^n-1$ elements since there are $2^n - 1$ distinct transpositions of $\ket{a}$ with another computational basis state.
\end{proof}

\section{Achieving nearly-optimal gate complexity for a computational basis state transposition}
\label{sect:achieve-exact}

In this section we present a quantum circuit construction to compile an arbitrary transposition. Our construction makes use of the $C^n X$ gate, so we first provide several statements on its decomposition into elementary gates. The main ideas behind these $C^n X $ decompositions can be traced back to \cite{Gidney2015}. In the following, we will refer to an ancilla qubit as a \textit{borrowed ancilla} if it can be in any initial state and its output state is unchanged. Similarly, we will refer to an ancilla qubit as a \textit{clean ancilla} if its initial and final state are both $\ket{0}$.

\begin{lemma}
\label{multiancillacnx}
    For all $n \geq 3$, a $C^n X $ gate can be compiled using $n-2$ borrowed ancilla qubits and at most $4n-8$ Toffoli gates.
\end{lemma}

The compilation and its proof are deferred to the Appendix, but we give the general construction now. We write $\Tof(i,j,k)$ to denote a Toffoli controlled on qubits $i,j$ and targeted on qubit $k$, and assume that a $C^n X $ is controlled on qubits $x_1,...,x_n$, targeting qubit $x_{n+1}$, and $a_1,...,a_{n-2}$ are borrowed ancillas. The sequence of Toffoli gates which implements the desired $C^n X$ operation is:
\begin{equation}
\label{toffolis1}
\Tof(a_{n-2}, x_n, x_{n+1}) \times \Big[ \Tof(a_{n-3}, x_{n-1}, a_{n-2}) \Tof(a_{n-4}, x_{n-2}, a_{n-3}) \dots \Tof(a_1, x_3, a_2) \Big] \times
\end{equation}
\begin{equation*}
\Big[ \Tof(x_1, x_2, a_1) \Tof(a_1, x_3, a_2) \dots \Tof(a_{n-4}, x_{n-2}, a_{n-3}) \Big] \times \Tof(a_{n-3}, x_{n-1}, a_{n-2})   
\end{equation*}
which is all repeated once more. The reader is directed to the Appendix for an explicitly worked out example of the above decomposition. The compilation of Lemma~\ref{multiancillacnx} uses a large number of ancilla qubits; However, this construction can be used for an alternative compilation with the same asymptotic gate complexity but which uses only a single clean ancilla. 

\begin{lemma}
\label{singleancillacnx}
    For all $n \geq 3$, a $C^n X$ gate can be compiled using one clean ancilla qubit and at most:
\begin{enumerate}[label=(\alph*)]
    \item $3$ Toffoli gates when $n=3$;
    \item $6n-18$ Toffoli gates for all $n \geq 4$.
\end{enumerate}
\end{lemma}

\begin{proof}
    Let $n_0 = \lceil n/2 \rceil$ and $n_1 = \lfloor n/2 \rfloor$ (thus $n_0 + n_1 = n$). We show that for all $n \geq 3$ the circuit:
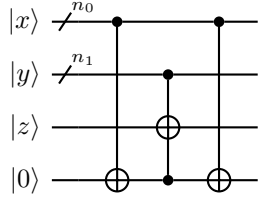
\begin{center}
\begin{quantikz}[column sep=10pt, row sep={20pt,between origins}]
            \lstick{$\ket{x}$} & \qwbundle{n_0} & \ctrl{3} & \qw       & \ctrl{3} & \qw \\
            \lstick{$\ket{y}$} & \qwbundle{n_1} & \qw      & \ctrl{1}  & \qw      & \qw \\
            \lstick{$\ket{z}$} & \qw            & \qw      & \targ{}   & \qw      & \qw \\
            \lstick{$\ket{0}$} & \qw            & \targ{}  & \ctrl{-1} & \targ{}  & \qw \\ 
\end{quantikz}    
\end{center}
acts as a $C^n X$ gate controlled on the first two registers and targeting the third, with the fourth register being a clean ancilla (where a control on a bundle of qubits represents a control on each qubit in the bundle). We prove this by showing that it implements the mapping which sends the basis state $\ket{x,y,z,0}$ to
\begin{equation*}
    \ket{x,y,z,0} \mapsto \ket{x,y,z \oplus (x_1 \wedge \dots \wedge x_{n_0}) \wedge (y_1 \wedge \dots \wedge y_{n_1}), 0}
\end{equation*}
for all $x=(x_1,...,x_{n_0}) \in \{0,1\}^{n_0}$, $y=(y_1,...,y_{n_1}) \in \{0,1\}^{n_1}$, and $z \in \{0,1\}$, where $\oplus$ denotes bit-wise addition and $\wedge$ denotes the logical ``and".

Considering the action of each operator in the circuit on an arbitrary initial state $\ket{x,y,z,0}$, the basis state is mapped as:
\begin{eqnarray}
&&\mapsto \ket{x,y,z,x_1 \wedge x_2 \wedge \cdots \wedge x_{n_0}} \nonumber \\
&&\mapsto |x,y,z \oplus (x_1 \wedge x_2 \wedge \cdots \wedge x_{n_0}) \wedge (y_1 \wedge \dots \wedge y_{n_1}), \nonumber \\
&& \hspace{0.5cm} x_1 \wedge x_2 \wedge \cdots \wedge x_{n_0} \rangle \nonumber \\
&&\mapsto \ket{x,y,z \oplus (x_1 \wedge x_2 \wedge \cdots \wedge x_{n_0}) \wedge (y_1 \wedge \dots \wedge y_{n_1}),0} \nonumber 
\end{eqnarray}
which shows the circuit implements the claimed operation.

Lastly, we count the number of Toffoli gates used. The circuit is composed of two $C^{n_0} X$ gates and one $C^{n_1+1} X$ gate. We compute the resulting Toffoli gate count by cases.

When $n=3$, then $n_0=2$ and $n_1=1$, so in this case we have used $3$ Toffoli gates and only one clean ancilla shown. This completes the proof of (a).

For (b) we first consider the case of $n=4$, where $n_0=2$ and $n_1=2$. In this case we may apply Lemma~\ref{multiancillacnx} to compile the $C^{n_1+1} X$ gate using one borrowed ancilla and $4$ Toffoli gates. There are $2$ qubits that are neither the target nor the control of the $C^{n_1+1} X$ gate and either may be used as a borrowed ancilla for its compilation. Therefore in this case we have used a total of $6$ Toffoli gates and only one clean ancilla, as claimed in (b), i.e., noting $6 \times 4 - 18 = 6$ Toffoli gates for $n=4$.


Finally, when $n\geq 5$ then both $n_0$ and $n_1+1$ are at least $3$ so we may compile the $C^{n_0} X$ and $C^{n_1+1} X$ gates using Lemma~\ref{multiancillacnx}. By Lemma~\ref{multiancillacnx}, a $C^{n_0} X$ gate can be compiled using $n_0-2$ borrowed ancilla qubits. Since $n_0 - 2 \leq n_1 + 1$, the $n_1+1$ qubits that are neither the target nor control of the $C^{n_0} X$ gates may be used as borrowed ancillas for their compilation. Similarly, a $C^{n_1+1} X$ gate can be compiled using $n_1-1$ borrowed ancilla qubits, and since $n_1-1 \leq n_0$, the $n_0$ qubits that are neither the target nor control of the $C^{n_1+1} X$ gate may be used as borrowed ancillas to compile it. Therefore no additional ancilla qubits are required. Counting Toffoli gates, we obtain a total of:
\begin{eqnarray}
2(4n_0 - 8) + 4(n_1+1) - 8 &&= 4n + 4n_0 - 20 \nonumber \\
&&\leq 4n + 2n + 2 - 20 \nonumber \\
&&= 6n - 18 \nonumber
\end{eqnarray}
gates, where the first equality follows since $n_0 + n_1 = n$, and the inequality follows since $4n_0 \leq 2n + 2$ (which holds because $n$ is an integer). Thus we have used the claimed number of Toffoli gates in (b). This completes the proof in all cases. 
\end{proof}

The final $C^n X$ compilation that we present uses a larger number of ancilla qubits, but reduces the number of Toffoli gates by a multiplicative constant. 

\begin{lemma}
\label{multicleanancillacnx}
    For all $n \geq 3$, a $C^n X$ gate can be compiled using $n-2$ clean ancilla qubits and $2n-3$ Toffoli gates.
\end{lemma}

\begin{proof}
    We provide a proof for the case $n=4$ for concreteness. The general case follows by an analogous argument on a circuit with the same pyramid-like shape that we present now. Consider the circuit:
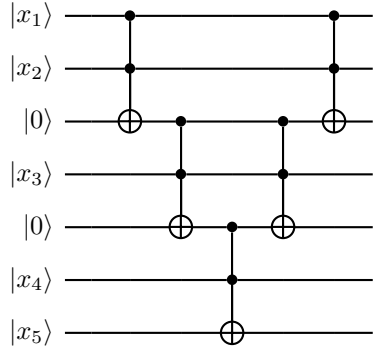
\begin{center}
\begin{quantikz}[column sep=10pt, row sep={20pt,between origins}]
            \lstick{$\ket{x_1}$} & \qw & \ctrl{2} & \qw       & \qw      & \qw      & \ctrl{2} & \qw \\
            \lstick{$\ket{x_2}$} & \qw & \ctrl{1} & \qw       & \qw      & \qw      & \ctrl{1} & \qw \\
            \lstick{$\ket{0}$}   & \qw & \targ{}  & \ctrl{2}  & \qw      & \ctrl{2} & \targ{}  & \qw \\
            \lstick{$\ket{x_3}$} & \qw & \qw      & \ctrl{1}  & \qw      & \ctrl{1} & \qw      & \qw \\ 
            \lstick{$\ket{0}$}   & \qw & \qw      & \targ{}   & \ctrl{2} & \targ{}  & \qw      & \qw \\
            \lstick{$\ket{x_4}$} & \qw & \qw      & \qw       & \ctrl{1} & \qw      & \qw      & \qw \\ 
            \lstick{$\ket{x_5}$} & \qw & \qw      & \qw       & \targ{}  & \qw      & \qw      & \qw \\ 
\end{quantikz}    
\end{center}   
We will show that this circuit acts as a $C^4 X$ gate which is controlled on the first, second, fourth, and sixth qubit, targets the final qubit, and the remaining qubits are clean ancillas. Applying the gates one at a time to an arbitrary initial computational basis state $\ket{x_1,x_2,0,x_3,0,x_4,x_5}$ it is mapped as:
\begin{eqnarray}
&&\mapsto \ket{x_1,x_2,x_1 \wedge x_2,x_3,0,x_4,x_5} \nonumber \\
&&\mapsto \ket{x_1,x_2,x_1 \wedge x_2,x_3,x_1 \wedge x_2 \wedge x_3,x_4,x_5} \nonumber \\
&&\mapsto |x_1,x_2,x_1 \wedge x_2,x_3,x_1 \wedge x_2 \wedge x_3,x_4, \nonumber \\
&& \hspace{0.5cm} x_5 \oplus (x_1 \wedge x_2 \wedge x_3 \wedge x_4) \rangle \nonumber \\
&&\mapsto \ket{x_1,x_2,x_1 \wedge x_2,x_3, 0 ,x_4,x_5 \oplus (x_1 \wedge x_2 \wedge x_3 \wedge x_4)} \nonumber \\
&&\mapsto \ket{x_1,x_2, 0,x_3, 0 ,x_4,x_5 \oplus (x_1 \wedge x_2 \wedge x_3 \wedge x_4)} \nonumber
\end{eqnarray}
which shows that the circuit implements the claimed operation. The number of Toffoli gates and clean ancillas follows by directly counting. 
\end{proof}

We can now give the main result of this section. To this end, let $\ket{a}$ and $\ket{b}$ be an arbitrary pair of $n$-qubit computational basis states that are to be transposed; further let $\Pi_a$ and $\Pi_b$ be projectors onto these basis states. Our construction will make use of the $(n+1)$-qubit operators:
\begin{equation}
\label{xacontrolledx}
    \Pi_a \otimes X + (I-\Pi_a) \otimes I,
\end{equation}
\begin{equation}
\label{xbcontrolledx}
    \Pi_b \otimes X + (I-\Pi_b) \otimes I.
\end{equation}
In circuit diagrams, these are represented as a block denoted $\Pi_a$ or $\Pi_b$ controlling a ``$\oplus$'' on the target qubit. As the projectors in question are onto computational basis states, these gates may be realised by a $C^n X$ gate where the control is ``sandwiched'' between a pair of $X$ gates when the conditioned on 0 for the relevant qubit. In this way, the gate ``picks out'' a single computational basis state which controls a bit flip on the target qubit. We also define the $n$-qubit operator:
\begin{equation*}
    U_{a,b}:= U_1 \otimes U_2 \otimes \dots \otimes U_n
\end{equation*}
where $U_i = X$ if $a$ and $b$ differ in the $i^{th}$ bit, and $U_i = I$ otherwise. Note that $U_{a,b}$ acts on $\ket{a}$ and $\ket{b}$ as $U_{a,b}\ket{a}=\ket{b}$ and $U_{a,b}\ket{b}=\ket{a}$.

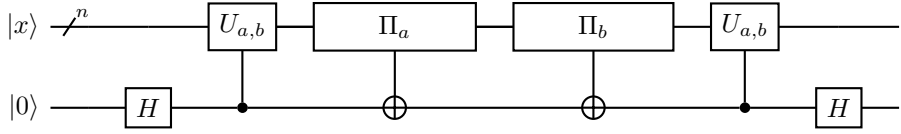
\begin{figure*}
\centering
\begin{quantikz}
            \lstick{$\ket{x}$} & \qwbundle{n} & \qw      & \gate{U_{a,b}} & \gate{\Pi_a}\ctrl{1} & \gate{\Pi_b}\ctrl{1} & \gate{U_{a,b}} & \qw      & \qw \\
            \lstick{$\ket{0}$} & \qw          & \gate{H} & \ctrl{-1}      & \targ{}              & \targ{}              & \ctrl{-1}      & \gate{H} & \qw \\ 
\end{quantikz}
\captionsetup{width=.9\linewidth}
\caption{The circuit of Theorem~\ref{transpositionthm} which acts as a computational basis state transposition.}
\label{fig:Transpositioncircuit}
\end{figure*}

\begin{thm}
\label{transpositionthm}
The circuit in Fig.~\ref{fig:Transpositioncircuit} acts as a transposition of the computational basis states $\ket{a}$ and $\ket{b}$ for all $n$. For $n=1$, $n=2$ and $n=3$ the circuit requires at most:
\begin{enumerate}[label=(\roman*)]
    \item $2$ Hadamard gates; $4$ X gates; $4$ CNOT gates; and one clean ancilla;
    \item $2$ Hadamard gates; $8$ X gates; $4$ CNOT gates; $2$ Toffoli gates; and one clean ancilla;
    \item $2$ Hadamard gates; $12$ X gates; $6$ CNOT gates; $6$ Toffoli gates; and 2 clean ancillas;
\end{enumerate}
respectively, and for all $n \geq 4$ requires at most either:
\begin{enumerate}[label=(\alph*)]
    \item $2$ Hadamard gates; $4n$ $X$ gates; $2n$ CNOT gates; $12n-36$ Toffoli gates; and $2$ clean ancillas; or
    \item $2$ Hadamard gates; $4n$ $X$ gates; $2n$ CNOT gates; $4n-6$ Toffoli gates; and $n-1$ clean ancillas. 
\end{enumerate}
Thus in all cases the overall gate complexity is $\Theta(n)$, nearly achieving the lower-bound of Corollary~\ref{cor1}.
\end{thm}
\begin{proof}
    First, we show that the circuit acts as the mapping defined in (\ref{transpositionmap}) for an arbitrary input $\ket{x}\ket{0}$:
    \begin{itemize}
        \item For $\ket{x}\ket{0}$ with $x \notin \{a,b\}$, the first Hadamard gate maps $\ket{x}\ket{0}$ to $\ket{x}\ket{+}$. As $U_{a,b}$ is a permutation which sends $\ket{a}$ to $\ket{b}$ and $\ket{b}$ to $\ket{a}$, it follows that $U_{a,b}$ must send $\ket{x}$ to some computational basis state $\ket{y}$, where $y$ is not equal to $a$ or $b$. Therefore the controlled-$U_{a,b}$ sends $\ket{x}\ket{+}$ to $\frac{1}{\sqrt{2}}\ket{x}\ket{0} + \frac{1}{\sqrt{2}}\ket{y}\ket{1}$. Following this, none of the conditions of the next two controlled operations are met, so the state remains unchanged. The state is then mapped by the last controlled-$U_{a,b}$ to $\frac{1}{\sqrt{2}}\ket{x}\ket{0} + \frac{1}{\sqrt{2}}\ket{x}\ket{1}$, and the final Hadamard gate maps this to $\ket{x}\ket{0}$. Therefore the overall operation in this case is to map $\ket{x}\ket{0}$ to $\ket{x}\ket{0}$.
        \item Turning to the case where $x=a$, the first Hadamard gate sends $\ket{a}\ket{0} \mapsto \frac{1}{\sqrt{2}}\ket{a}\ket{0} + \frac{1}{\sqrt{2}}\ket{a}\ket{1}$; the controlled-$U_{a,b}$ operation then sends this to $\frac{1}{\sqrt{2}}\ket{a}\ket{0} + \frac{1}{\sqrt{2}}\ket{b}\ket{1}$; the third and fourth circuit block together send this to $\frac{1}{\sqrt{2}}\ket{a}\ket{1} + \frac{1}{\sqrt{2}}\ket{b}\ket{0}$; the second controlled-$U_{a,b}$ operation then sends this to $\frac{1}{\sqrt{2}}\ket{b}\ket{1} + \frac{1}{\sqrt{2}}\ket{b}\ket{0}$ and the remaining Hadamard gate maps this to $\ket{b}\ket{0}$. Thus the overall operation is to send $\ket{a}\ket{0} \mapsto \ket{b}\ket{0}$. The case where $x=b$ follows by a completely analogous argument.
    \end{itemize}
    By the above case analysis, we have shown that the circuit performs the claimed transposition.

    Having shown that the circuit has the required operation, it remains to count gates and qubits. There are two uses of the controlled-$U_{a,b}$ operator. Each requires at most $n$ CNOT gates, giving at most $2n$ CNOT gates. Continuing, there are two uses of the operators defined in (\ref{xacontrolledx}) and (\ref{xbcontrolledx}). Each of these operators consists of a $C^n X$ gate and at most $2n$ additional $X$ gates. Thus for any $n \geq 1$ the total number of operations required is: $2$ Hadamard gates; $4n$ X gates; $2n$ CNOT gates; and 2 $C^n X$ gates. For the cases $n=1$ and $n=2$ this results in the bounds claimed in (i) and (ii). For $n\geq 3$, we may apply either Lemma~\ref{singleancillacnx} or \ref{multicleanancillacnx} to compile each $C^n X$ (and it turns out that for the case of $n=3$ the resources are identical). Using the compilation provided by Lemma~\ref{singleancillacnx}, each $C^n X$ can be compiled with a clean ancilla qubit and $3$ Toffoli gates when $n=3$, or $6n-18$ Toffoli gates when $n \geq 4$. Since the required ancilla qubit is a clean ancilla, we may reuse the same one for each of these operations. Therefore in this case we have used a total of: two clean ancillas -- one explicitly shown and one required by Lemma~\ref{singleancillacnx}; $6$ Toffoli gates for $n=3$ and $12n-36$ Toffoli gates for $n\geq 4$; $2n$ CNOT gates; $2$ Hadamard gates; and at most $4n$ $X$ gates. This completes the proof of (a). For (b), suppose we instead apply the $C^n X$ compilation of Lemma~\ref{multicleanancillacnx}. In this case each $C^n X$ can be compiled using $n-2$ clean ancillas and $2n-3$ Toffoli gates. Since the ancillas are clean then they may be reused for each of these operations. Therefore the total gate complexity in this case is: $n-1$ clean ancillas -- one explicitly shown and $n-2$ required by Lemma~\ref{multicleanancillacnx}; $4n-6$ Toffoli gates; $2n$ CNOT gates; $2$ Hadamard gates; and at most $4n$ $X$ gates -- completing the proof of (b).
\end{proof}

\begin{remark}
    The number of $X$ gates can be reduced to $3n$ by noticing that for any qubit controlled on the 0 state for both the $\Pi_a$ and $\Pi_b$ controlled $X$ gates, a pair of $X$ gates will cancel. That is, owing to the fact that these gates occur consecutively, there will be three rather than four (partially filled) banks of $X$ gates, when the trivial simplification $XX = I$ is applied to the compilation. We also note the similarity of the circuit structure in Fig.~\ref{fig:Transpositioncircuit} to that of the Hadamard test, which may be useful for further generalizations in future work. 
\end{remark}

\section{Numerical Results}
\label{sect:numericalresults}

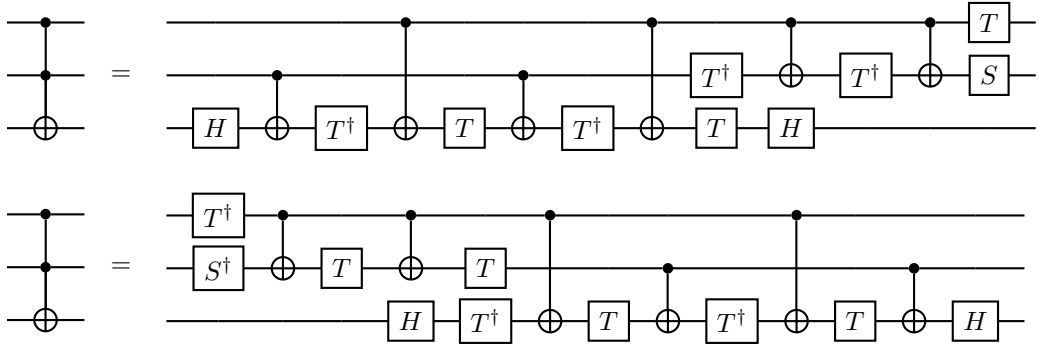
\begin{figure*}
\begin{subfigure}{1.0 \textwidth}
\begin{quantikz}[column sep=10pt, row sep={20pt,between origins}]
    \qw & \ctrl{2} & \qw & \ghost{H}  & \ghost{H}  \\
    \qw & \ctrl{1} & \qw & =          & \ghost{H}  \\
    \qw & \targ{}  & \qw & \ghost{H}  & \ghost{H}  \\
\end{quantikz}  \begin{quantikz}[column sep=10pt, row sep={20pt,between origins}]
            \qw & \qw      & \qw       & \qw              & \ctrl{2} & \qw      & \qw      & \qw              & \ctrl{2} & \qw              & \ctrl{1}  & \qw  
            & \ctrl{1} & \gate{T} & \qw \\
            \qw & \qw      & \ctrl{1}  & \qw              & \qw      & \qw      & \ctrl{1} & \qw              & \qw      & \gate{T^\dagger} & \targ{}   & \gate{T^\dagger} & \targ{}  & \gate{S} & \qw \\
            \qw & \gate{H} & \targ{}   & \gate{T^\dagger} & \targ{}  & \gate{T} & \targ{}  & \gate{T^\dagger} & \targ{}  & \gate{T}         & \gate{H}  & \qw              & \qw      & \qw      & \qw \\
\end{quantikz}
\end{subfigure}%
\hfill
\begin{subfigure}{1.0 \textwidth}
\begin{quantikz}[column sep=10pt, row sep={20pt,between origins}]
    \qw & \ctrl{2} & \qw & \ghost{H}  & \ghost{H}  \\
    \qw & \ctrl{1} & \qw & =          & \ghost{H}  \\
    \qw & \targ{}  & \qw & \ghost{H}  & \ghost{H}  \\
\end{quantikz}  \begin{quantikz}[column sep=10pt, row sep={20pt,between origins}]
            \qw & \gate{T^\dagger} & \ctrl{1} & \qw      & \ctrl{1} & \qw              & \ctrl{2} & \qw      & \qw       & \qw              & \ctrl{2} & \qw      & \qw      & \qw      & \qw \\
            \qw & \gate{S^\dagger} & \targ{}  & \gate{T} & \targ{}  & \gate{T}         & \qw      & \qw      & \ctrl{1}  & \qw              & \qw      & \qw      & \ctrl{1} & \qw      & \qw \\
            \qw & \qw              & \qw      & \qw      & \gate{H} & \gate{T^\dagger} & \targ{}  & \gate{T} & \targ{}   & \gate{T^\dagger} & \targ{}  & \gate{T} & \targ{}  & \gate{H} & \qw \\
\end{quantikz}
\end{subfigure}
\captionsetup{width=.9\linewidth}
\caption{The standard decomposition of the Toffoli gate (above) and its inverse (below) into single-qubit and CNOT gates.\cite[Fig.~4.9]{nielsenchuang2010}}
\label{fig:Toffolidecomp}
\end{figure*}

We now present some numerical results to demonstrate the performance of our proposed method of transposing computational basis states.  The numerical results fall into two categories. First, we compile a range of transpositions using the approach described in Theorem~\ref{transpositionthm} (a) and (b) and compare the CNOT and Toffoli gate counts of the resulting circuits to the theoretical bounds described therein. Second, we compare the CNOT and T gate counts of our method against several state-of-the-art approaches for compiling permutational circuits, namely the Tweedledum-based construction presented in \cite{Soeken2019} and the ToffoliBox of pytket. In each case, the Toffoli gates are compiled according to Fig.~\ref{fig:Toffolidecomp}, such that the final circuits consist only of CNOT and single-qubit gates. The choice to compare CNOT and T gate counts is motivated by the fact that typically CNOT gates are the most expensive gates when running circuits with physical qubits, and T gates are the most expensive gates to perform fault-tolerantly. All of the resulting circuits were compiled using pytket 1.18.0, and only mild optimisation passes were used to simplify gate redundancies. See \href{https://cqcl.github.io/tket/pytket/api/}{https://cqcl.github.io/tket/pytket/api/} for pytket documentation.

\begin{table}[h]
\centering
\caption{\label{tab:CX-Toff-counts}%
The average CNOT and Toffoli gate counts across $200$ random transpositions. The number $n$ is the number of qubits required for the computational basis states that are transposed, and ``Bd" is an abbreviation for the word ``Bound". `(a)' and `(b)' refer to the two settings considered in Theorem~\ref{transpositionthm}; note that the bound on the number of CNOTS is the same for both (a) and (b).
}
\begin{tabular}{ccccccccc}
&\multicolumn{4}{c}{CNOT}&\multicolumn{4}{c}{Toffoli} \\
n & Avg (a) & Avg (b) & Bd & n & Avg (a) & Bd (a) & Avg (b) & Bd (b)\\
\hline
$2$ & $2.60$ & $2.64$ & $4$ & $2$ & $2$ & $2$ & $2$ & $2$ \\
$3$ & $3.52$ & $3.35$ & $6$ & $3$ & $6$ & $6$ & $6$ & $6$ \\
$4$ & $4.10$ & $4.18$ & $8$ & $4$ & $12$ & $12$ & $10$ & $10$ \\
$5$ & $5.13$ & $5.15$ & $10$ & $5$ & $24$ & $24$ & $14$ & $14$ \\
$6$ & $6.10$ & $6.10$ & $12$ & $6$ & $32$ & $36$ & $18$ & $18$ \\
$7$ & $7.12$ & $6.95$ & $14$ & $7$ & $48$ & $48$ & $22$ & $22$ \\
$8$ & $8.33$ & $8.05$ & $16$ & $8$ & $56$ & $60$ & $26$ & $26$ \\
$9$ & $8.87$ & $9.00$ & $18$ & $9$ & $72$ & $72$ & $30$ & $30$ \\
$10$ & $10.09$ & $10.36$ & $20$ & $10$ & $80$ & $84$ & $34$ & $34$ \\
$11$ & $11.14$ & $10.75$ & $22$ & $11$ & $96$ & $96$ & $38$ & $38$ \\
$12$ & $11.95$ & $12.30$ & $24$ & $12$ & $104$ & $108$ & $42$ & $42$ \\
$13$ & $12.66$ & $13.09$ & $26$ & $13$ & $120$ & $120$ & $46$ & $46$ \\
$14$ & $14.05$ & $14.05$ & $28$ & $14$ & $128$ & $132$ & $50$ & $50$ \\
$15$ & $14.78$ & $15.02$ & $30$ & $15$ &  $144$ & $144$ & $54$ & $54$ \\
$16$ & $15.86$ & $15.82$ & $32$ & $16$ &  $152$ & $156$ & $58$ & $58$ \\
$17$ & $17.03$ & $16.55$ & $34$ & $17$ &  $168$ & $168$ & $62$ & $62$ \\
$18$ & $18.43$ & $17.64$ & $36$ & $18$ &  $176$ & $180$ & $66$ & $66$ \\
$19$ & $18.39$ & $19.24$ & $38$ & $19$ &  $192$ & $192$ & $70$ & $70$ \\
$20$ & $20.12$ & $20.74$ & $40$ & $20$ &  $200$ & $204$ & $74$ & $74$ \\
\end{tabular}
\end{table}

\subsection{Comparison with theoretical bounds}

For our first set of results, we compare the average CNOT and Toffoli gate counts across a range of random transpositions to the theoretical bounds described in Theorem~\ref{transpositionthm}. For each $2 \leq n \leq 20$, we generate $200$ random transpositions of two $n$-qubit computational basis states and use the constructions proposed in Theorem~\ref{transpositionthm} (a) and (b) to compile their corresponding circuits, resulting in circuits over the gate-set $\{H, X, \textnormal{CNOT}, \textnormal{Toffoli}\}$. The RemoveRedundacies pass in pytket was applied to each of these circuits and then the average CNOT and Toffoli gate counts were tabulated and presented in Tables~\ref{tab:CX-Toff-counts}. The results show that the average CNOT count is typically only approximately half of the bound that we prove in this paper; whereas the Toffoli counts saturate or nearly saturate the bounds in all cases.

\subsection{Comparison with other approaches}

For our second set of results we compare the average CNOT and T gate counts of the compilation method proposed in Theorem~\ref{transpositionthm} to the Tweedledum-based compilation method of Ref. \cite{Soeken2019} and the ToffoliBox of pytket. In each case either $100$ transpositions of the same Hamming weight were randomly generated, or in the case where there are fewer than $100$ distinct tranpositions of the same Hamming weight then the entire set of them was considered. In particular, in the following cases there were fewer than 100 possible transpositions (format is (number of qubits, Hamming distance) total transpositions): (4, 1) 32;
(4, 2) 48;
(4, 3) 32;
(4, 4) 8;
(5, 1) 80;
(5, 4) 80;
(5, 5) 16;
(6, 6) 32;
(7, 7) 64. For each Hamming weight the resulting circuits were compiled and the average CNOT count and T gate counts were computed and presented in Figures~\ref{fig:CXcount} and \ref{fig:Tcount}.  

\textit{Tweedledum:} The first compilation method that we compare our method to is that of Ref. \cite{Soeken2019}. For each of the random transpositions the circuits were compiled and simplified using the CliffordSimp, SynthesiseTket and RemoveReduncies passes in pytket, which resulted in circuits using CNOT, TK1, and global phase gates. Pytket 1.18.0 does not contain functionality for T gate synthesis, and so only the CNOT gate counts were recorded and presented in Fig.~\ref{fig:CXcount}.

\textit{ToffoliBox:} The second compilation method is the ToffoliBox of pytket. The compilation can use one of two strategies, referred to as ``Matching" and ``Cycle". For the matching strategy, the resulting circuits were compiled and simplified using the CliffordSimp, SynthesiseTket and RemoveReduncies passes in pytket, resulting in circuits that use CNOT, TK1, and global phase gates. As noted in the Tweedledum case, pytket 1.18.0 does not contain functionality for T gate synthesis, and so only the CNOT gate counts were recorded and presented in Fig.~\ref{fig:CXcount} as Pytket-Match. It is worth mentioning that compilation from a gate-set including the continuously-parameterised gate TK1 to a finite gate-set, such as that containing the Clifford gates and T gates (or Cliffords, Toffolis and T gates), can only be done approximately, and if very high accuracy is required, the T gate count becomes large. For this reason, fault-tolerant compilation is likely to favor techniques that only require gates from a suitable finite set in the first place.

For the cycle strategy, the ToffoliBox returns a circuit consisting of $X$ and $C^n X $ gates. To more readily compare these circuits to those of our proposed construction, the $C^n X $ gates were decomposed into $X$, CNOT and Toffoli gates using the same $C^n X$ decompositions used in Theorem~\ref{transpositionthm} (a) and (b). The Toffoli gates were then decomposed into single-qubit gates and CNOT gates using the standard decomposition of Fig.~\ref{fig:Toffolidecomp}. The RemoveRedundancies pass of pytket was then applied and the CNOT and T gate counts were recorded. The counts are denoted by Pytket-Cycle (a) and (b) in Figures~\ref{fig:CXcount} and \ref{fig:Tcount} corresponding to the $C^n X$ decomposition used. 

\textit{Theorem~\ref{transpositionthm}:} For the compilation method of Theorem~\ref{transpositionthm} the circuits were compiled into $X$, CNOT and Toffoli gates using the constructions described in the theorem. Following this, the Toffoli gates in the circuits were then decomposed into single-qubit gates and CNOT gates using the decomposition of Fig.~\ref{fig:Toffolidecomp}. The RemoveRedundancies pass of pytket was then applied and the CNOT and T gate counts were recorded. The corresponding counts are denoted by Thm~\ref{transpositionthm} (a) and Thm~\ref{transpositionthm} (b) in Figures~\ref{fig:CXcount} and \ref{fig:Tcount}. 

We can see that the methods we propose in this paper are relatively most advantageous for large numbers of qubits and for large Hamming distances. This is to be expected, as our methods are nearly optimal in the number of qubits, and have approximately the same performance for any transposition -- whereas other methods, such as those that use a Gray code, suffer when the transposition is such that the Hamming distance between transposed computational basis states (written as binary strings) is large.

\begin{figure}[h]
\centering
\begin{subfigure}{.5 \textwidth}
\centering
\begin{tikzpicture}[scale=0.70]
\begin{axis}[
    title={},
    xlabel={Number of qubits},
    ylabel={Average CNOT count},
    xmin=4, xmax=8,
    ymin=0, ymax=800,
    xtick={4,5,6,7,8},
    ytick={200,400,600,800},
    legend pos=north west,
    ymajorgrids=true,
    grid style=dashed,
]

\addplot[color=blue,mark=o]
    coordinates {(4,14)(5,30)(6,62)(7,126)(8,254)};
    \addlegendentry{Tweedledum}
\addplot[color=red,mark=o]
    coordinates {(4,20.1)(5,51.1)(6,127.2)(7,287.2)(8,623.2)};    
    \addlegendentry{Pytket-Match}
\addplot[color=red,mark=square]
    coordinates {(4,18)(5,36)(6,72)(7,96)(8,144)};    
    \addlegendentry{Pytket-Cycle (a)}
\addplot[color=red,mark=triangle]
    coordinates {(4,18)(5,30)(6,42)(7,54)(8,66)};    
    \addlegendentry{Pytket-Cycle (b)}
\addplot[color=black,mark=square]
    coordinates {(4,68)(5,146)(6,194)(7,290)(8,338)};    
    \addlegendentry{Thm 3 (a)}
\addplot[color=black,mark=triangle]
    coordinates {(4,53)(5,71.6)(6,90.8)(7,105.8)(8,126.9)};    
    \addlegendentry{Thm 3 (b)} 
\end{axis}
\end{tikzpicture} 
\caption{Hamming distance $1$}
\end{subfigure}%
\begin{subfigure}{.5 \textwidth}
\centering
\begin{tikzpicture}[scale=0.70]
\begin{axis}[
    title={},
    xlabel={Number of qubits},
    ylabel={Average CNOT count},
    xmin=4, xmax=8,
    ymin=0, ymax=1200,
    xtick={4,5,6,7,8},
    ytick={200,400,600,800,1000,1200},
    legend pos=north west,
    ymajorgrids=true,
    grid style=dashed,
]

\addplot[color=blue,mark=o]
    coordinates {(4,41.5)(5,89.6)(6,185.6)(7,377.7)(8,761.8)};
    \addlegendentry{Tweedledum}
\addplot[color=red,mark=o]
    coordinates {(4,45.8)(5,106.7)(6,226.7)(7,504.7)(8,1107.1)};    
    \addlegendentry{Pytket-Match}
\addplot[color=red,mark=square]
    coordinates {(4,50)(5,99.8)(6,216)(7,288)(8,432)};    
    \addlegendentry{Pytket-Cycle (a)}
\addplot[color=red,mark=triangle]
    coordinates {(4,50)(5,78.2)(6,112.6)(7,142.8)(8,163.9)};    
    \addlegendentry{Pytket-Cycle (b)}
\addplot[color=black,mark=square]
    coordinates {(4,74)(5,148)(6,196)(7,292)(8,340)};    
    \addlegendentry{Thm 3 (a)}
\addplot[color=black,mark=triangle]
    coordinates {(4,62)(5,82.4)(6,105.2)(7,126.4)(8,143.0)};    
    \addlegendentry{Thm 3 (b)}
    
\end{axis}
\end{tikzpicture}
\caption{Hamming distance $2$}    
\end{subfigure}%
\\
\begin{subfigure}{.5 \textwidth}
\centering
\begin{tikzpicture}[scale=0.70]
\begin{axis}[
    title={},
    xlabel={Number of qubits},
    ylabel={Average CNOT count},
    xmin=4, xmax=8,
    ymin=0, ymax=1500,
    xtick={4,5,6,7,8},
    ytick={300,600,900,1200,1500},
    legend pos=north west,
    ymajorgrids=true,
    grid style=dashed,
]

\addplot[color=blue,mark=o]
    coordinates {(4,68.8)(5,149.1)(6,309)(7,629.5)(8,1269.5)};
    \addlegendentry{Tweedledum}
\addplot[color=red,mark=o]
    coordinates {(4,58)(5,133.4)(6,293.4)(7,628.1)(8,1347.7)};    
    \addlegendentry{Pytket-Match}
\addplot[color=red,mark=square]
    coordinates {(4,78)(5,161.0)(6,360)(7,480)(8,720)};    
    \addlegendentry{Pytket-Cycle (a)}
\addplot[color=red,mark=triangle]
    coordinates {(4,78)(5,124.8)(6,169.4)(7,214.1)(8,257.3)};    
    \addlegendentry{Pytket-Cycle (b)}
\addplot[color=black,mark=square]
    coordinates {(4,78)(5,150)(6,198)(7,294)(8,342)};    
    \addlegendentry{Thm 3 (a)}
\addplot[color=black,mark=triangle]
    coordinates {(4,66)(5,89.2)(6,111.5)(7,133.8)(8,154.9)};    
    \addlegendentry{Thm 3 (b)}
    
\end{axis}
\end{tikzpicture}
\caption{Hamming distance $3$}    
\end{subfigure}%
\begin{subfigure}{0.5\textwidth}
\centering
\begin{tikzpicture}[scale=0.70]
\begin{axis}[
    title={},
    xlabel={Number of qubits},
    ylabel={Average CNOT count},
    xmin=4, xmax=8,
    ymin=0, ymax=2000,
    xtick={4,5,6,7,8},
    ytick={400,800,1200,1600,2000},
    legend pos=north west,
    ymajorgrids=true,
    grid style=dashed,
]

\addplot[color=blue,mark=o]
    coordinates {(4,96)(5,208.3)(6,432.7)(7,881.2)(8,1777.3)};
    \addlegendentry{Tweedledum}
\addplot[color=red,mark=o]
    coordinates {(4,69)(5,153.4)(6,339.6)(7,750.8)(8,1602.4)};    
    \addlegendentry{Pytket-Match}
\addplot[color=red,mark=square]
    coordinates {(4,102)(5,218.4)(6,504)(7,672)(8,1008)};    
    \addlegendentry{Pytket-Cycle (a)}
\addplot[color=red,mark=triangle]
    coordinates {(4,102)(5,162)(6,223)(7,276)(8,338.6)};    
    \addlegendentry{Pytket-Cycle (b)}
\addplot[color=black,mark=square]
    coordinates {(4,80)(5,152)(6,200)(7,296)(8,344)};    
    \addlegendentry{Thm 3 (a)}
\addplot[color=black,mark=triangle]
    coordinates {(4,68)(5,92)(6,115.2)(7,137.5)(8,160.9)};    
    \addlegendentry{Thm 3 (b)}
    
\end{axis}
\end{tikzpicture}
\caption{Hamming distance $4$}
\end{subfigure}
\\
\begin{subfigure}{0.5\textwidth}
\centering
\begin{tikzpicture}[scale=0.70]
\begin{axis}[
    title={},
    xlabel={Number of qubits},
    ylabel={Average CNOT count},
    xmin=4, xmax=8,
    ymin=0, ymax=2400,
    xtick={4,5,6,7,8},
    ytick={600,1200,1800,2400},
    legend pos=north west,
    ymajorgrids=true,
    grid style=dashed,
]

\addplot[color=blue,mark=o]
    coordinates {(5,268)(6,556.6)(7,1132.9)(8,2285)};
    \addlegendentry{Tweedledum}
\addplot[color=red,mark=o]
    coordinates {(5,173)(6,375.5)(7,819.9)(8,1778.7)};    
    \addlegendentry{Pytket-Match}
\addplot[color=red,mark=square]
    coordinates {(5,276)(6,648)(7,864)(8,1296)};    
    \addlegendentry{Pytket-Cycle (a)}
\addplot[color=red,mark=triangle]
    coordinates {(5,198)(6,273.1)(7,344.6)(8,417.4)};    
    \addlegendentry{Pytket-Cycle (b)}
\addplot[color=black,mark=square]
    coordinates {(5,154)(6,202)(7,298)(8,346)};    
    \addlegendentry{Thm 3 (a)}
\addplot[color=black,mark=triangle]
    coordinates {(5,94)(6,118)(7,141.6)(8,164.4)};    
    \addlegendentry{Thm 3 (b)}
    
\end{axis}
\end{tikzpicture}
\caption{Hamming distance $5$}
\end{subfigure}%
\begin{subfigure}{0.5\textwidth}
\centering
\begin{tikzpicture}[scale=0.70]
\begin{axis}[
    title={},
    xlabel={Number of qubits},
    ylabel={Average CNOT count},
    xmin=4, xmax=8,
    ymin=0, ymax=3000,
    xtick={4,5,6,7,8},
    ytick={700,1400,2100,2800},
    legend pos=north west,
    ymajorgrids=true,
    grid style=dashed,
]

\addplot[color=blue,mark=o]
    coordinates {(6,680)(7,1384.5)(8,2792.7)};
    \addlegendentry{Tweedledum}
\addplot[color=red,mark=o]
    coordinates {(6,413)(7,896.2)(8,1951.7)};    
    \addlegendentry{Pytket-Match}
\addplot[color=red,mark=square]
    coordinates {(6,792)(7,1056)(8,1584)};    
    \addlegendentry{Pytket-Cycle (a)}
\addplot[color=red,mark=triangle]
    coordinates {(6,318)(7,403.2)(8,477.8)};    
    \addlegendentry{Pytket-Cycle (b)}
\addplot[color=black,mark=square]
    coordinates {(6,204)(7,300)(8,348)};    
    \addlegendentry{Thm 3 (a)}
\addplot[color=black,mark=triangle]
    coordinates {(6,120)(7,144)(8,167.4)};    
    \addlegendentry{Thm 3 (b)}
    
\end{axis}
\end{tikzpicture}
\caption{Hamming distance $6$}
\end{subfigure}%
\captionsetup{width=.9\linewidth}
\caption{The average CNOT counts across $100$ randomly selected transpositions (or over all transpositions, when the total is fewer than 100) between computational basis states with a fixed Hamming distance. The number of qubits is the number of qubits required for the computational basis states that are transposed.}
\label{fig:CXcount}
\end{figure}

\begin{figure}[h]
\centering
\begin{subfigure}{0.5\textwidth}
\centering
\begin{tikzpicture}[scale=0.70]
\begin{axis}[
    title={},
    xlabel={Number of qubits},
    ylabel={Average T gate count},
    xmin=4, xmax=8,
    ymin=0, ymax=400,
    xtick={4,5,6,7,8},
    ytick={50,100,150,200,250,300,350,400},
    legend pos=north west,
    ymajorgrids=true,
    grid style=dashed,
]
\addplot[color=red,mark=square]
    coordinates {(4,21)(5,42)(6,84)(7,112)(8,168)};    
    \addlegendentry{Pytket-Cycle (a)}
\addplot[color=red,mark=triangle]
    coordinates {(4,21)(5,35)(6,49)(7,63)(8,77)};    
    \addlegendentry{Pytket-Cycle (b)}
\addplot[color=black,mark=square]
    coordinates {(4,77)(5,168)(6,224)(7,336)(8,392)};    
    \addlegendentry{Thm 3 (a)}
\addplot[color=black,mark=triangle]
    coordinates {(4,59.5)(5,81.2)(6,103.6)(7,121.1)(8,145.7)};    
    \addlegendentry{Thm 3 (b)}
\end{axis}
\end{tikzpicture}
\caption{Hamming distance $1$}
\end{subfigure}%
\begin{subfigure}{0.5\textwidth}
\centering
\begin{tikzpicture}[scale=0.70]
\begin{axis}[
    title={},
    xlabel={Number of qubits},
    ylabel={Average T gate count},
    xmin=4, xmax=8,
    ymin=0, ymax=600,
    xtick={4,5,6,7,8},
    ytick={100,200,300,400,500,600},
    legend pos=north west,
    ymajorgrids=true,
    grid style=dashed,
]
\addplot[color=red,mark=square]
    coordinates {(4,58.3)(5,116.5)(6,252)(7,336)(8,504)};    
    \addlegendentry{Pytket-Cycle (a)}
\addplot[color=red,mark=triangle]
    coordinates {(4,58.3)(5,91.8)(6,131.3)(7,166.6)(8,191.2)};    
    \addlegendentry{Pytket-Cycle (b)}
\addplot[color=black,mark=square]
    coordinates {(4,81.7)(5,168)(6,224)(7,336)(8,392)};    
    \addlegendentry{Thm 3 (a)}
\addplot[color=black,mark=triangle]
    coordinates {(4,67.7)(5,91.4)(6,118.2)(7,142.8)(8,162.1)};    
    \addlegendentry{Thm 3 (b)}
\end{axis}
\end{tikzpicture}
\caption{Hamming distance $2$}    
\end{subfigure}%
\\
\begin{subfigure}{0.5\textwidth}
\centering
\begin{tikzpicture}[scale=0.70]
\begin{axis}[
    title={},
    xlabel={Number of qubits},
    ylabel={Average T gate count},
    xmin=4, xmax=8,
    ymin=0, ymax=1000,
    xtick={4,5,6,7,8},
    ytick={200,400,600,800,1000},
    legend pos=north west,
    ymajorgrids=true,
    grid style=dashed,
]
\addplot[color=red,mark=square]
    coordinates {(4,91)(5,187.9)(6,420)(7,560)(8,840)};    
    \addlegendentry{Pytket-Cycle (a)}
\addplot[color=red,mark=triangle]
    coordinates {(4,91)(5,145.6)(6,197.7)(7,249.8)(8,300.2)};    
    \addlegendentry{Pytket-Cycle (b)}
\addplot[color=black,mark=square]
    coordinates {(4,84)(5,168)(6,224)(7,336)(8,392)};    
    \addlegendentry{Thm 3 (a)}
\addplot[color=black,mark=triangle]
    coordinates {(4,70)(5,97)(6,123.1)(7,149.1)(8,173.8)};    
    \addlegendentry{Thm 3 (b)}
\end{axis}
\end{tikzpicture}
\caption{Hamming distance $3$}    
\end{subfigure}%
\begin{subfigure}{0.5\textwidth}
\centering
\begin{tikzpicture}[scale=0.70]
\begin{axis}[
    title={},
    xlabel={Number of qubits},
    ylabel={Average T gate count},
    xmin=4, xmax=8,
    ymin=0, ymax=1200,
    xtick={4,5,6,7,8},
    ytick={200,400,600,800,1000,1200},
    legend pos=north west,
    ymajorgrids=true,
    grid style=dashed,
]
\addplot[color=red,mark=square]
    coordinates {(4,119)(5,254.8)(6,588)(7,784)(8,1176)};    
    \addlegendentry{Pytket-Cycle (a)}
\addplot[color=red,mark=triangle]
    coordinates {(4,119)(5,189)(6,260.1)(7,322)(8,395.1)};    
    \addlegendentry{Pytket-Cycle (b)}
\addplot[color=black,mark=square]
    coordinates {(4,84)(5,168)(6,224)(7,336)(8,392)};    
    \addlegendentry{Thm 3 (a)}
\addplot[color=black,mark=triangle]
    coordinates {(4,70)(5,98)(6,125)(7,151.1)(8,178.4)};    
    \addlegendentry{Thm 3 (b)}
\end{axis}
\end{tikzpicture}
\caption{Hamming distance $4$}    
\end{subfigure}%
\\
\begin{subfigure}{0.5\textwidth}
\centering
\begin{tikzpicture}[scale=0.70]
\begin{axis}[
    title={},
    xlabel={Number of qubits},
    ylabel={Average T gate count},
    xmin=4, xmax=8,
    ymin=0, ymax=1600,
    xtick={4,5,6,7,8},
    ytick={400,800,1200,1600},
    legend pos=north west,
    ymajorgrids=true,
    grid style=dashed,
]
\addplot[color=red,mark=square]
    coordinates {(5,322)(6,756)(7,1008)(8,1512)};    
    \addlegendentry{Pytket-Cycle (a)}
\addplot[color=red,mark=triangle]
    coordinates {(5,231)(6,318.6)(7,402.1)(8,486.9)};    
    \addlegendentry{Pytket-Cycle (b)}
\addplot[color=black,mark=square]
    coordinates {(5,168)(6,224)(7,336)(8,392)};    
    \addlegendentry{Thm 3 (a)}
\addplot[color=black,mark=triangle]
    coordinates {(5,98)(6,126)(7,153.6)(8,180.2)};    
    \addlegendentry{Thm 3 (b)}
\end{axis}
\end{tikzpicture}
\caption{Hamming distance $5$}    
\end{subfigure}%
\begin{subfigure}{0.5\textwidth}
\centering
\begin{tikzpicture}[scale=0.70]
\begin{axis}[
    title={},
    xlabel={Number of qubits},
    ylabel={Average T gate count},
    xmin=4, xmax=8,
    ymin=0, ymax=2000,
    xtick={4,5,6,7,8},
    ytick={400,800,1200,1600,2000},
    legend pos=north west,
    ymajorgrids=true,
    grid style=dashed,
]
\addplot[color=red,mark=square]
    coordinates {(6,924)(7,1232)(8,1848)};    
    \addlegendentry{Pytket-Cycle (a)}
\addplot[color=red,mark=triangle]
    coordinates {(6,371)(7,470.4)(8,557.5)};    
    \addlegendentry{Pytket-Cycle (b)}
\addplot[color=black,mark=square]
    coordinates {(6,224)(7,336)(8,392)};    
    \addlegendentry{Thm 3 (a)}
\addplot[color=black,mark=triangle]
    coordinates {(6,126)(7,154)(8,181.3)};    
    \addlegendentry{Thm 3 (b)}
\end{axis}
\end{tikzpicture}
\caption{Hamming distance $6$}    
\end{subfigure}%
\captionsetup{width=.9\linewidth}
\caption{The average T gate counts across $100$ randomly selected transpositions (or over all transpositions, when the total is fewer than 100) between computational basis states with a fixed Hamming distance. The number of qubits is the number of qubits required for the computational basis states that are transposed.}
\label{fig:Tcount}
\end{figure}

\begin{figure}

\label{fig:randomtranspositionToffolicount}
\end{figure}

\section{Discussion}
\label{sect:discussion}

In this paper we have shown that on average $n$-qubit computational basis states transpositions have a gate complexity $\Omega(n / \log(nd))$ when using any $d$-element gate-set, and even if ancillas are available. Since a general permutation can be expressed as a product of at most $2^{n-1}$ transpositions then this lower bound is consistent with the $\Omega(n2^n/\log(n))$ worst-case lower bound of Ref~\cite{Shende2003} for an arbitrary permutation. We subsequently give an explicit construction to perform any computational basis state transposition with $\Theta(n)$ gates and two ancillas. To our knowledge, this is first time that this construction has been proposed, and conventional wisdom is to use the Gray code construction popularised in Nielsen and Chuang \cite[Section~4.5.2]{nielsenchuang2010} to perform any 2-level unitary (in the case of a transposition the unitary is the Pauli-$X$ matrix), which requires $\Theta(n^2)$ gates in the worst case. This therefore represents a potentially practically useful result for any compiler that constructs arbitrary permutations from transpositions. This claim of potential for practical utility is backed-up by the numerical results presented, which show that for transpositions with large numbers of qubits and / or large Hamming distance, our methods outperform the standard alternatives.

It is also worth noting that the transposition construction presented in Theorem~\ref{transpositionthm} is amenable to several further circuit optimisations during compilation. In particular, if we consider the compilation of a Toffoli gate into single-qubit gates and CNOTs, then the standard circuit is given by Fig.~\ref{fig:Toffolidecomp}. However, as the Toffoli gate is equal to its inverse, we also have that the circuit reversed and with every gate replaced by its inverse also implements the Toffoli, as shown in Fig.~\ref{fig:Toffolidecomp}. Therefore it follows that, each time a pair of Toffoli gates appear as:
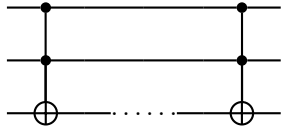
\begin{center}
\begin{quantikz}[column sep=10pt, row sep={20pt,between origins}]
    \qw & \ctrl{2} & \qw & \qw   & \qw & \ctrl{2} & \qw \\
    \qw & \ctrl{1} & \qw & \qw   & \qw & \ctrl{1} & \qw \\
    \qw & \targ{}  & \qw & \push{\dots \dots} & \qw & \targ{}  & \qw \\
\end{quantikz}
\end{center}

(where the dotted line implies that other operations occur here) then we can use the second Toffoli decomposition for the second Toffoli, such that the decomposed circuit is:
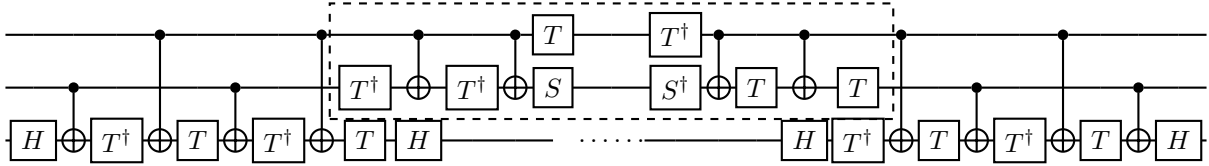
\begin{center}
\begin{quantikz}[column sep=2pt, row sep={20pt,between origins}]
            \qw & \qw      & \qw       & \qw              & \ctrl{2} & \qw      & \qw      & \qw              & \ctrl{2}  & \qw \gategroup[2,steps=11,style={dashed, inner sep=0.1pt}]{}              & \ctrl{1}  & \qw              & \ctrl{1} & \gate{T} & \qw         & \gate{T^\dagger} & \ctrl{1} & \qw      & \ctrl{1} & \qw              & \ctrl{2} & \qw      & \qw       & \qw              & \ctrl{2} & \qw      & \qw      & \qw      & \qw \\            
            \qw & \qw      & \ctrl{1}  & \qw              & \qw      & \qw      & \ctrl{1} & \qw              & \qw      & \gate{T^\dagger} & \targ{}   & \gate{T^\dagger} & \targ{}  & \gate{S} & \qw         & \gate{S^\dagger} & \targ{}  & \gate{T} & \targ{}  & \gate{T}         & \qw      & \qw      & \ctrl{1}  & \qw              & \qw      & \qw      & \ctrl{1} & \qw      & \qw \\            
            \qw & \gate{H} & \targ{}   & \gate{T^\dagger} & \targ{}  & \gate{T} & \targ{}  & \gate{T^\dagger} & \targ{}  & \gate{T}         & \gate{H}  & \qw              & \qw      & \qw      & \dots \dots & \qw              & \qw      & \qw      & \gate{H} & \gate{T^\dagger} & \targ{}  & \gate{T} & \targ{}   & \gate{T^\dagger} & \targ{}  & \gate{T} & \targ{}  & \gate{H} & \qw
\end{quantikz}    
\end{center}
where we can readily see that the gates inside of the region enclosed by the dashed line cancel to the identity. So it follows that we have implemented the two Toffolis using a total of 8 CNOTs and 12 single-qubit gates -- fewer than the 12 CNOTs and 20 single-qubit gates that are needed in general to compile two Toffolis. We can further see, for example in (\ref{cnxcircuit}), such structures are commonplace in our construction, and hence has the potential for significant CNOT and T gate count reductions during compilation. These savings can be readily observed in the numerical data presented in Figures~\ref{fig:CXcount} and \ref{fig:Tcount}, which demonstrates lower CNOT and T gate counts for transpositions between computational basis states of large Hamming distance when compared to Tweedledum and Pytket.

\section*{Acknowledgements}

The authors would like to thank Silas Dilkes, Alexandre Krajenbrink, and Tuomas Laakkonen for carefully reviewing and providing useful feedback on an earlier draft of this article. Special thanks to Tuomas for suggesting improvements to the circuit construction given in Theorem~\ref{transpositionthm}, and to Silas for providing various suggestions for Section~\ref{sect:numericalresults}.

\bibliography{mybib}{}
\bibliographystyle{IEEEtran}

\appendix
\section*{Appendix}

We provide an explicit example of Lemma~\ref{multiancillacnx} when $n=4$, as described by (\ref{toffolis1}). The circuit in this case is
\begin{equation}
\label{cnxcircuit}
\begin{quantikz}[column sep=10pt, row sep={20pt,between origins}]
            \lstick{$\ket{x_1}$} & \qw      & \qw      & \ctrl{2}  & \qw      & \qw      & \qw      & \ctrl{2} & \qw      \\
            \lstick{$\ket{x_2}$} & \qw      & \qw      & \ctrl{1}  & \qw      & \qw      & \qw      & \ctrl{1} & \qw      \\
            \lstick{$\ket{a_1}$} & \qw      & \ctrl{2} & \targ{}   & \ctrl{2} & \qw      & \ctrl{2} & \targ{}  & \ctrl{2} \\
            \lstick{$\ket{x_3}$} & \qw      & \ctrl{1} & \qw       & \ctrl{1} & \qw      & \ctrl{1} & \qw      & \ctrl{1} \\
            \lstick{$\ket{a_2}$} & \ctrl{2} & \targ{}  & \qw       & \targ{}  & \ctrl{2} & \targ{}  & \qw      & \targ{}  \\
            \lstick{$\ket{x_4}$} & \ctrl{1} & \qw      & \qw       & \qw      & \ctrl{1} & \qw      & \qw      & \qw      \\
            \lstick{$\ket{x_5}$} & \targ{}  & \qw      & \qw       & \qw      & \targ{}  & \qw      & \qw      & \qw      \\
\end{quantikz}
\end{equation}
which we show implements the desired mapping. Note that the general case follows by considering a circuit with the same pyramid structure as above, and where qubits number $2i+3$, for $i=0,1,...,n-3$, are ancilla qubits. We consider the action of the circuit in (\ref{cnxcircuit}) on an arbitrary input $\ket{x_1,x_2,a_1,x_3,a_2,x_4,x_5}$ and show that it implements the mapping:
\begin{equation}
\label{cnxmap}
    \ket{x_1,x_2,a_1,x_3,a_2,x_4,x_5} \mapsto \ket{x_1,x_2,a_1,x_3,a_2,x_4,x_5 \oplus (x_1 \wedge \dots \wedge x_4)}
\end{equation}

\textbf{Case 1:} Suppose that at least one of $x_1$ or $x_2$ is equal to $0$. Then the two Toffoli gates that are controlled on the first two registers will act as the identity. Consequentially, the circuit in (\ref{cnxcircuit}) will have the same action as the circuit:
\begin{equation}
\begin{quantikz}[column sep=10pt, row sep={20pt,between origins}]\label{circ1}
            \lstick{$\ket{x_1}$} & \qw      & \qw        & \qw      & \qw      & \qw       & \qw      \\
            \lstick{$\ket{x_2}$} & \qw      & \qw        & \qw      & \qw      & \qw       & \qw      \\
            \lstick{$\ket{a_1}$} & \qw      & \ctrl{2}   & \ctrl{2} & \qw      & \ctrl{2}  & \ctrl{2} \\
            \lstick{$\ket{x_3}$} & \qw      & \ctrl{1}   & \ctrl{1} & \qw      & \ctrl{1}  & \ctrl{1} \\
            \lstick{$\ket{a_2}$} & \ctrl{2} & \targ{}    & \targ{}  & \ctrl{2} & \targ{}   & \targ{}  \\
            \lstick{$\ket{x_4}$} & \ctrl{1} & \qw        & \qw      & \ctrl{1} & \qw       & \qw      \\
            \lstick{$\ket{x_5}$} & \targ{}  & \qw        & \qw      & \targ{}  & \qw       & \qw      \\
\end{quantikz}    
\end{equation}
Since the Toffoli gate is equal to its inverse then the consecutive Toffoli gates multiply to the identity, and so (\ref{circ1}) is equivalent to:
\begin{equation}
\begin{quantikz}[column sep=10pt, row sep={20pt,between origins}]
            \lstick{$\ket{x_1}$} & \qw      & \qw      \\
            \lstick{$\ket{x_2}$} & \qw      & \qw      \\
            \lstick{$\ket{a_1}$} & \qw      & \qw      \\
            \lstick{$\ket{x_3}$} & \qw      & \qw      \\
            \lstick{$\ket{a_2}$} & \ctrl{2} & \ctrl{2} \\
            \lstick{$\ket{x_4}$} & \ctrl{1} & \ctrl{1} \\
            \lstick{$\ket{x_5}$} & \targ{}  & \targ{}  \\
\end{quantikz}    
\end{equation}
which is then equivalent to the empty circuit for the same reason. Therefore the overall mapping is to send:
\begin{equation}
\label{identitymap}
    \ket{x_1,x_2,a_1,x_3,a_2,x_4,x_5} \mapsto \ket{x_1,x_2,a_1,x_3,a_2,x_4,x_5}
\end{equation}
in this case.

\textbf{Case 2:} Suppose that $x_3$ is equal to $0$. Then the four Toffoli gates which target the fifth qubit of (\ref{cnxcircuit}) act as the identity, and the circuit acts as:
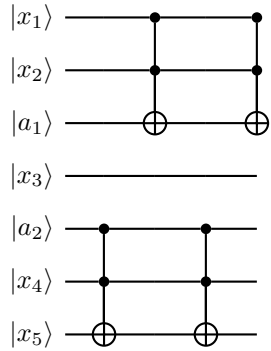
\begin{center}
\begin{quantikz}[column sep=10pt, row sep={20pt,between origins}]
            \lstick{$\ket{x_1}$} & \qw      & \ctrl{2}  & \qw      & \ctrl{2} \\
            \lstick{$\ket{x_2}$} & \qw      & \ctrl{1}  & \qw      & \ctrl{1} \\
            \lstick{$\ket{a_1}$} & \qw      & \targ{}   & \qw      & \targ{}  \\
            \lstick{$\ket{x_3}$} & \qw      & \qw       & \qw      & \qw      \\
            \lstick{$\ket{a_2}$} & \ctrl{2} & \qw       & \ctrl{2} & \qw      \\
            \lstick{$\ket{x_4}$} & \ctrl{1} & \qw       & \ctrl{1} & \qw      \\
            \lstick{$\ket{x_5}$} & \targ{}  & \qw       & \targ{}  & \qw      \\
\end{quantikz}    
\end{center}
Again as the Toffoli gate is equal to its inverse then the above circuit acts as the identity, and the overall mapping in this case is also (\ref{identitymap}).

\textbf{Case 3:} Suppose that $x_4$ is equal to $0$. Then the two Toffoli gates that are targeted on the final qubit have no effect, and the circuit acts as:
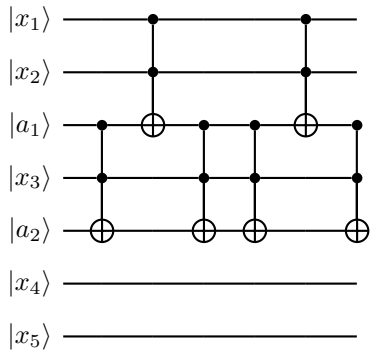
\begin{center}
\begin{quantikz}[column sep=10pt, row sep={20pt,between origins}]
            \lstick{$\ket{x_1}$} & \qw      & \ctrl{2}  & \qw      & \qw      & \ctrl{2} & \qw      \\
            \lstick{$\ket{x_2}$} & \qw      & \ctrl{1}  & \qw      & \qw      & \ctrl{1} & \qw      \\
            \lstick{$\ket{a_1}$} & \ctrl{2} & \targ{}   & \ctrl{2} & \ctrl{2} & \targ{}  & \ctrl{2} \\
            \lstick{$\ket{x_3}$} & \ctrl{1} & \qw       & \ctrl{1} & \ctrl{1} & \qw      & \ctrl{1} \\
            \lstick{$\ket{a_2}$} & \targ{}  & \qw       & \targ{}  & \targ{}  & \qw      & \targ{}  \\
            \lstick{$\ket{x_4}$} & \qw      & \qw       & \qw      & \qw      & \qw      & \qw      \\
            \lstick{$\ket{x_5}$} & \qw      & \qw       & \qw      & \qw      & \qw      & \qw      \\
\end{quantikz}    
\end{center}
By the same reasoning as in the previous cases, the circuit reduces to the identity.

\textbf{Case 4:} Lastly, we check the operation when $x_1,x_2,x_3,x_4$ are all equal to $1$. Considering the circuit (\ref{cnxcircuit}) one gate at a time the basis state $\ket{1, 1, a_1, 1, a_2, 1, x_5}$ is mapped to:
\begin{eqnarray*}
&&\mapsto \ket{1, 1, a_1,          1, a_2,            1, x_5 \oplus a_2 }  \\
&&\mapsto \ket{1, 1, a_1,          1, a_2 \oplus a_1, 1, x_5 \oplus a_2 }  \\
&&\mapsto \ket{1, 1, a_1 \oplus 1, 1, a_2 \oplus a_1, 1, x_5 \oplus a_2 }  \\
&&\mapsto \ket{1, 1, a_1 \oplus 1, 1, a_2 \oplus 1,   1, x_5 \oplus a_2 }  \\
&&\mapsto \ket{1, 1, a_1 \oplus 1, 1, a_2 \oplus 1,   1, x_5 \oplus 1 }    \\
&&\mapsto \ket{1, 1, a_1 \oplus 1, 1, a_2 \oplus a_1, 1, x_5 \oplus 1 }    \\
&&\mapsto \ket{1, 1, a_1,          1, a_2 \oplus a_1, 1, x_5 \oplus 1 }    \\
&&\mapsto \ket{1, 1, a_1,          1, a_2,            1, x_5 \oplus 1 }    
\end{eqnarray*}
which indeed is the operation in (\ref{cnxmap}). Therefore we checked all cases and shown that the circuit in (\ref{cnxcircuit}) implements the mapping (\ref{cnxmap}), as claimed.

\end{document}